\documentstyle[12pt]{article}
\newcommand{\beq}{\begin{equation}}
\newcommand{\eeq}{\end{equation}}
\newcommand{\bea}{\begin{eqnarray}}
\newcommand{\eea}{\end{eqnarray}}
\newcommand{\x}{{\bf x}}

\newcommand{\k}{\kappa}
\newcommand{\e}{\epsilon}

\newcommand{\eb}{{\overline e}}

\newcommand{\wb}{{\overline \omega}}

\newcommand{\p}{\phi}
\newcommand{\w}{\omega}

\newcommand{\T}{{\cal T}}
\renewcommand{\d}{\delta}
\renewcommand{\l}{\lambda}
\renewcommand{\L}{\Lambda}
\renewcommand{\b}{\beta}
\renewcommand{\a}{\alpha}
\newcommand{\n}{\nu}
\newcommand{\m}{\mu}
\newcommand{\r}{\rho}
\newcommand{\s}{\sigma}

\newcommand{\oh}{\frac{1}{2}}
\newcommand{\oq}{\frac{1}{4}}

\newcommand{\non}{\nonumber}

\renewcommand{\t}{\tau}
\newcommand{\rf}[1]{(\ref{#1})}
\newcommand{\ra}{\rightarrow}
\begin{document}
\topmargin 0pt
\oddsidemargin 5mm
\headheight 0pt
\headsep 0pt
\topskip 9mm

\addtolength{\baselineskip}{0.20\baselineskip}
\hfill    NBI-HE-92-41

\hfill June 1992
\begin{center}

\vspace{36pt}
{\large \bf VANISHING OF THE COSMOLOGICAL CONSTANT IN
STABILIZED QUANTUM GRAVITY}\footnote{Supported by the
U.S. Department of Energy under Grant No. DE-FG03-92ER40711.}

\end{center}

\vspace{36pt}

\begin{center}
{\sl J. Greensite}
\footnote{Permanent address: Physics and Astronomy Dept.,
San Francisco State University, San Francisco, CA 94132 USA.} \\

\vspace{12pt}

The Niels Bohr Institute\\
Blegdamsvej 17\\
DK-2100 Copenhagen \O , Denmark\\

\end{center}

\vfill

\begin{center}
{\bf Abstract}
\end{center}

\vspace{12pt}

   It is shown that the probability distribution $P(\l)$ for the
effective cosmological constant is sharply peaked at $\l=0$
in stochastic (or "fifth-time") stabilized quantum gravity.  The
effect is similar to the Baum-Hawking mechanism, except that it
comes about due to quantum fluctuations, rather than as a zeroth-order
(in $\hbar$) semiclassical effect.

\vspace{24pt}

\vfill

\newpage

   Very simple and elegant arguments for the vanishing of the
cosmological constant were advanced some time ago, in the context of Euclidean
quantum gravity, by Baum \cite{Baum}, Hawking \cite{Hawking}, and
Coleman \cite{Coleman}.  The Baum-Hawking argument, in particular,
can be summarized in a few lines:  The effective cosmological constant
$\l$ is the sum of the bare cosmological constant $\l_0$ plus a
contribution from the stress-energy of the non-gravitational fields, and
it is assumed that the range of possible $\l$ includes the value $\l=0$.
Integrating over the gravitational field then gives a probability
distribution for $\l$,

\beq
       P(\l) = \int Dg \; exp \left[-\int d^4x \sqrt{g}
[-{1\over 16\pi G}R + \l] \right]
\label{Pdef}
\eeq
and the semiclassical evaluation of this expression gives

\beq
       P(\l) \approx exp[{3 \over 8 G^2 \l}]
\label{BH}
\eeq
which is peaked at infinity as $\l \rightarrow 0^+$.  This infinite peak
in the probability distribution at $\l=0^+$ is the Baum-Hawking
explanation for the vanishing of the cosmological constant.  Coleman's
argument \cite{Coleman}, which appeals to wormhole effects, is a related
idea, and gives at the semiclassical level a double exponential
distribution

\beq
       P(\l) \approx exp[exp[{3 \over 8 G^2 \l}]]
\label{Cman}
\eeq

   Unfortunately there is a strong objection to the
Baum-Hawking-Coleman arguments; namely, that these arguments rely on
(and suffer from) the fact that the Euclidean Einstein-Hilbert action is
unbounded from below.  At the classical level, the action at the stationary
point $\d S =0$ is $S=-3/(8G^2\l)$, which leads to the semiclassical
distributions \rf{BH} and \rf{Cman} above, but in fact the action can be
made arbitrarily negative, due to the well-known "wrong-sign" of the
kinetic term of the metric conformal factor.  This unboundedness of
the action from below means that the functional integral in
\rf{Pdef}, and the Euclidean quantum theory based on it, are
essentially meaningless.  If one nevertheless attempts to define the integral
over the conformal factor by a contour rotation \cite{Gibbons},
then the unboundedness problem is simply shifted to the matter
Lagrangian \cite{Horowitz}.  Even in pure gravity, the conformal
rotation introduces a complex phase $(-i)^{D+2}$
in front of the second exponential of eq. \rf{Cman}, which ruins the
Coleman argument in D=4 dimensions \cite{Polchinski}.  Finally,
deformation of a field integration contour into the complex plane
tends to generate complex expectation
values of physical quantities at the non-perturbative level; this is
known to occur in matrix models for 2D gravity \cite{David}.

   However, there exists a general method, grounded in stochastic quantization,
for defining the Euclidean quantum theory of any action unbounded from
below.  This is the stochastic stabilization, or "fifth-time action"
method, introduced in ref. \cite{GH}, which has the property of stabilizing
the action while preserving the classical limit and, if the bottomless
action is stable at zeroth order, also preserving the naive perturbative
expansion to all orders in any coupling constant.  The method has
been applied to D=0 matrix models representing 2D gravity in ref.
\cite{matmod}, and to the Einstein-Hilbert and Einstein-Cartan
actions of 4D gravity in ref. \cite{Me1} and \cite{Me2}.  Very recently
there has also been some been work on the cosmological constant issue in
stochastic stabilized quantum gravity.  Carlini and Martellini
\cite{CM} have studied a minisuperspace approximation of the stabilized
Einstein-Hilbert action, while the author, in ref. \cite{Me2}, has carried
out a Monte Carlo calculation of the stabilized, latticized,
Einstein-Cartan theory.  In both cases there appears to be evidence
for a peaking of the probability distribution around $\l=0^+$.  These
results motivate the one-loop calculation presented here.

  It was first noted by Giveon et. al. \cite{Giveon} (reviewed in
more detail in ref. \cite{Me2}), that the fifth-time action method of
ref. \cite{GH} is completely equivalent to Langevin evolution, i.e.
stochastic quantization, between fixed non-singular (but otherwise
arbitrary) initial and final field configurations in the fictitious
time.  The precise prescription for this "stochastic stabilization"
is as follows:
Let $g^M$ be the fields, $G_{MN}$ the supermetric, and $E^A_M$ the
supervielbein, of a given field theory.  Evolution in the fictitious
time $t_5$ is given by the Langevin equation

\beq
    \partial_5 g^M(\x,t_5) = -G^{MN} {\d S \over \d g^N} + E^M_A \eta^A
\label{Langevin}
\eeq
starting from an initial configuration at $t_5=-T$

\beq
     g^M(\x,-T) = g^M_i(\x)
\eeq
Then the expectation value $<Q>$ of any operator $Q[g^M(\x)]$ in the
stabilized Euclidean theory is given by

\bea
     <Q> &=& \lim_{T \ra \infty} {1 \over Z_5} \int
D\eta(\x,-T<t_5<T) \; Q[g^M(\x,0)] \d[g^M(\x,T) - g^M_f(\x)]
\nonumber \\
    & & \times exp[-\int^T_{-T} d^5x \; \eta^A \eta_A / 4 \hbar]
\label{stable}
\eea

    The only difference between the prescription \rf{stable}, and
ordinary stochastic quantization, is the delta function enforcing
the final state constraint $g^M(\x,T)=g^M_f(\x)$.  This constraint
prevents the system from running away to a singular configuration
as $T \ra \infty$, thereby stabilizing the theory.
It can be shown in general that in the
$T \ra \infty$ limit, $<Q>$ is independent of the choice of initial and
final configurations $g^M_i$ and $g^M_f$.\footnote{In particular, for
ordinary bounded actions, the prescription \rf{stable} is equivalent to the
usual formulations.} By a change of variables (c.f. \cite{Me2}),
eq. \rf{stable} can be converted into the form

\bea
     <Q> &=& {1 \over Z_5} \int Dg^M(\x,t_5) \; \sqrt{G} \; Q[g^M(\x,0)]
e^{-S_5/\hbar}
\nonumber \\
      S_5 &=& \int d^5x [\oq G_{MN} \partial_5 g^M \partial_5 g^N
+ \oq G^{MN} {\d S \over \d g^M} {\d S \over \d g^N} -
{\hbar \over 2} G^{MN} {\d^2 S \over \d g^M \d g^N} ]
\label{S_5}
\eea
where $S_5$ is the "fifth-time" action.

   Applying stochastic stabilization to Einstein-Cartan gravity

\beq
      S_{EC} = \int \e_{abcd}[- {1 \over 4 \k^2} e^a \land e^b \land
(d \w^{cd} +\w^{cf}\land \w^{fd}) + \l_0 \int d^4x \; det(e)
\label{SEC}
\eeq
leads to the stabilized formulation

\beq
     <Q[e,\w]> = {1 \over Z_5} \int De D\w \; \sqrt{G}
\; Q[e(\x,0),\w(\x,0)] e^{-S_5/\hbar}
\label{A}
\eeq
where
\bea
   S_5 &=& \oq \int d^5x \sqrt{g} \left[
{1 \over \k^2} g^{\m\n} (\partial_5 e^a_\m \partial_5 e^a_\n + \hbar
\partial_5 \w^{ab}_\m \partial_5 \w^{ab}_\n) \right.
\non \\
& & + 4({1\over \k^2}R^{a}_{\m} R^{a}_{\n} g^{\m\n}
-\l_0 R + \k^2 \l_0^2)
\non \\
 & & \left. + {1 \over  \k^4 \hbar} T^a_{\m \n} T^b_{\r \s}
(\d_{ab} g^{\m\r} + 2 e^\m_a e^\r_b)g^{\n\s} \right]
\label{S5EC}
\eea
is the 5-th time action and

\bea
       R &=& d\w + \w \wedge \w
\non \\
       T &=& de + \w \wedge e
\eea
are the curvature and torsion two-forms respectively. The root
determinant of the $e-\w$ supermetric is

\beq
      \sqrt{G} = \sqrt{G_e} \sqrt{G_\w} \propto
\prod_{\x,t_5} det^{10}(e)
\eeq
where $G_e$ and $G_\w$ denote the separate tetrad and spin-connection
supermetrics.  The range of functional integration is
restricted to tetrads representing compact 4-manifolds, at each $t_5$.

   The stabilized theory of Euclidean gravity, defined above, is
invariant with respect to D=4 diffeomorphisms, yields the
usual Einstein field equations in its classical limit, contains
no higher-derivative terms, and appears
to be reflection-positive (at least in certain lattice versions) across
the ordinary time ($x_4$) axis \cite{Me1}.  It can also be
expressed in terms of an effective four-dimensional action

\bea
    \lefteqn{ exp\{-S_{eff}[e'(\x),\w'(\x)]/\hbar\} }
\non \\
&=& \int De(\x,t_5) D\w(\x,t_5)
\; \sqrt{G} \; \d[e(\x,0)-e'(\x)] \d[\w(\x,0)-\w'(\x)] e^{-S_5/\hbar}
\eea
which can be calculated perturbatively, as discussed in \cite{Me1}.

   We now consider coupling stabilized gravity to non-gravitational
fields, denoted collectively by $\p$.  Treating $\p(\x)$ as a
$t_5$-independent source leads to the stabilized four-dimensional
action

\bea
    \lefteqn{ exp\{-S_{eff}[\p(\x),e'(\x),\w'(\x)]/\hbar\} }
\non \\
&=& \int De(\x,t_5) D\w(\x,t_5)
\; \sqrt{G} \; \d[e(\x,0)-e'(\x)] \d[\w(\x,0)-\w'(\x)] e^{-S_5/\hbar}
\eea
and expectation values

\beq
     <Q> = {1 \over Z} \int D\p(\x) \int De(\x,t_5) D\w(\x,t_5) \;
 \sqrt{G} \;  Q[\p(\x),e(\x,0),\w(\x,0)] e^{-S_5/\hbar}
\label{stable2}
\eeq
where
\bea
   S_5 &=& \oq \int d^5x \sqrt{g} \left[
{1 \over \k^2} g^{\m\n} (\partial_5 e^a_\m \partial_5 e^a_\n + \hbar
\partial_5 \w^{ab}_\m \partial_5 \w^{ab}_\n) \right.
\non \\
& & + {\k^2 \over 4}g_{\a\b}
[{2 \over \k^2}(R^{\a}_c - \oh R e^{\a}_c) + \l_0 e^\a_c + \T^\a_c]
[{2 \over \k^2}(R^{\b}_c - \oh R e^{\b}_c) + \l_0 e^\b_c + \T^\b_c]
\non \\
 & & \left. + {1 \over  \k^4 \hbar} T^a_{\m \n} T^b_{\r \s}
(\d_{ab} g^{\m\r} + 2 e^\m_a e^\r_b)g^{\n\s} \right]
\label{S5T}
\eea
and $\T_{\m\n}(\p(\x))$ is the stress-energy tensor for the non-gravitational
$\p$ fields.

     In principle one could include the $\p$ fields, along
with the tetrad and spin-connection, among the $t_5$-dependent $g^M$ fields
of the Langevin equation \rf{Langevin}.  This is not done here for
two reasons: first, in the classical $\hbar \ra 0$ limit, \rf{S5T}
implies the Einstein field equations,

\beq
      R_{\m\n} - \oh R g_{\m\n} + {\k^2 \over 2} \l_0 g_{\m\n} =
          -{\k^2 \over 2} \T_{\m \n}
\label{Einstein}
\eeq
and the Einstein field equations in turn imply the classical equations
of motion of the $\p$ fields \cite{MTW}.  Stabilizing the (already
stable)
non-gravitational fields via the Langevin equation seems
redundant.  Second and more importantly, if the $\p$ are dynamic in $t_5$,
then the fifth-time action $S_5$ contains higher-derivative terms
coming from $(\d S / \d \p)^2$, which threaten reflection-positivity for
reflections across $x_4$.  This problem is avoided if the
non-gravitational fields are treated as $t_5$-independent, as above.

  From \rf{stable2} we see that integration over the tetrad $e=\eb +\d e$
and spin connection $\w = \wb + \d \w$ produces a probability measure
for the non-gravitational fields on a ($\p$-dependent) background manifold
$(\eb,\wb)$, i.e.

\beq
        P[\p(\x)] = {1 \over Z} e^{-S_{eff}[\p]/\hbar} = {1 \over Z}
 \int De(\x,t_5) D\w(\x,t_5) \; \sqrt{G} \; e^{-S_5[e,\w,\p]/\hbar}
\eeq
Let us consider, in particular, non-gravitational field configurations,
such as \linebreak
$\phi=const.$, in which
the stress-energy is proportional to the metric tensor

\beq
        \T_{\m\n} = \t g_{\m\n}
\eeq
Then the stress-energy can be combined with $\l_0$ to produce an
effective cosmological constant

\beq
       \l = \t + \l_0
\eeq
with a probability distribution

\beq
        P[\l] = {1 \over Z}  \int De(\x,t_5) D\w(\x,t_5) \;
\sqrt{G}  \; e^{-S_5[e,\w,\l]/\hbar}
\label{Pl}
\eeq
where $S_5[e,\w,\l]$ is identical to the $S_5$ in eq. \rf{S5EC},
with the replacement $\l_0 \ra \l$.

   The action in \rf{Pl} is bounded from below, and it is now
meaningful to ask whether the probability $P(\l)$ is peaked as
$\l \ra 0^+$.  For a given $\l>0$, the integral is to be evaluated
by expansion around

\bea
      e^a_\m(\x,t_5) &=& \eb^a_\m(\x) + \d e^a_\m(\x,t_5)
\non \\
      \w^{ab}_\m(\x,t_5) &=& \wb^{ab}_\m(e) + \k^2 \sqrt{\hbar}
\Omega^{ab}_\m(\x,t_5)
\eea
where $\eb(\x)$ is a tetrad for the 4-sphere of radius

\beq
       r = ({6 \over \k^2 \l})^{1/2}
\eeq
and $\wb(e)$ is the zero-torsion spin-connection

\beq
      \wb^{ab}_\m = \oh e^{\n a}(\partial_\m e^b_\n - \partial_\n
e^b_\m ) - \oh e^{\n b}(\partial_\m e^a_\n - \partial_\n e^a_\m )
- \oh e^{\r a}e^{\s b}(\partial_\r e_{\s c} - \partial_\s e_{\r c})
e^c_\m
\eeq
Then the part of $S_5$ which is zeroth-order in $\hbar$ is

\bea
   S_5 &=& \oq \int d^5x \sqrt{g} \left[ {1 \over \k^2} g^{\m\n}
\partial_5 e^a_\m \partial_5 e^a_\n  + 4({1\over \k^2}{\overline R}^a_\m
{\overline R}^a_\s g^{\m\s} - \l {\overline R} + \k^2 \l^2 ) \right.
\non \\
& & \left. + (\Omega \wedge e)^a_{\m\n} (\Omega \wedge e)^b_{\r \s} g^{\n\s}
(\d_{ab} g^{\m\r} + 2 e^\m_a e^\r_b) + O(\sqrt{\hbar}) \right]
\label{S5EC0}
\eea
where ${\overline R} = d\wb + \wb \wedge \wb$.  Integrating over
$\Omega$ cancels a factor $\sqrt{G_\w}$ in the measure, leaving

\bea
        P[\l] &=& {1 \over Z} \int De(\x,t_5)  \; \sqrt{G_e} \;
 e^{-S_5[e,\l]/\hbar}
\non \\
   S_5[e,\l] &=& \oq \int d^5x \sqrt{g} \left[ {1 \over \k^2} g^{\m\n}
\partial_5 e^a_\m \partial_5 e^a_\n  + 4({1\over \k^2}{\overline R}^a_\m
{\overline R}^a_\s g^{\m\s} - \l {\overline R} + \k^2 \l^2 ) \right]
\label{Pl1}
\eea

   The Baum-Hawking mechanism is not immediately apparent in \rf{Pl1},
simply because $S_5$ is bounded, and in fact vanishes when evaluated at
the classical solution $S_5[\eb,\l]=0$.  So if
there is a peak in $P[\l]$ at $\l=0^+$, it would have to come from quantum
fluctuations.  Fortunately this is just what happens, as we now show
at one-loop level.  Quantum gravity is perturbatively
non-renormalizable; this situation is unchanged by stabilization, so to
do a perturbative evaluation of \rf{Pl1} we must introduce a
short-distance cutoff $\L^{-1}$, presumably on the order of the Planck
length, where some new physics or fundamental granularity of spacetime
comes into play.  Expanding the action in \rf{Pl1} only up to terms
quadratic in the fluctuations

\beq
      S_5 \approx \int d^5x \; \d e M[\eb] \d e
\eeq
and integrating over the fluctuations yields the one-loop result

\beq
       P(\l) \approx det^{-1/2}[M]
\eeq
where the determinant needs to be regulated by, e.g., the heat-kernel
technique.  As is well-known, the regulated determinant can be
expanded in powers of the background Riemann tensor \cite{DeWitt,Gilkey}

\beq
       P(\l) \approx exp[c_0 \int d^5x \sqrt{{\overline g}}
+ c_1 \int d^5x \sqrt{{\overline g}} R({\overline g}) + ...]
\eeq
where $c_0,c_1$ are cutoff-dependent constants.  Because of the
non-renormalizability of quantum gravity we are not entitled to
drop such constants and just concentrate on the cutoff-independent
terms.  The cutoff at the Planck length is a true physical scale
where new physics is encountered, perhaps in the form of strings or
a fundamental lattice.  The largest powers of $\L$ and $1/\l$
are in the first term of the expansion, and $c_0$ is most easily
calculated by setting $\l=0$, and calculating the determinant around
flat space.  Expanding $S_5$ around flat space, $e_{\m a} = \d_{\m a} +
\d e_{\m a}$, we find for the part quadratic in fluctuations

\bea
    S_5 &=& {1 \over \k^2} \int d^5x \;
\d e_{\m\n}\left[ \oq(-\partial_5^2)\d_{\m\a}
\d_{\n\b} + (\partial_\m \partial^\m)^2 \{P^{(2)} + 4P^{(0s)}\}_{\m\n\a\b}
  \right] \d e_{\a\b}
\nonumber \\
     &=& {1 \over \k^2} \int d^5x \; \d e_{\m\n} \left\{
   [\oq(-\partial_5^2) +  (\partial_\m \partial^\m)^2] P^{(2)}
 + [\oq(-\partial_5^2) + 4(\partial_\m \partial^\m)^2] P^{(0s)} \right.
\nonumber \\
& & \left.
 +\oq(-\partial_5^2)({\bf 1}_A + P^{(1)} + P^{(0w)}) \right\}_{\m\n\a\b}
\d e_{\a\b}
\eea
where
\beq
{\bf 1}_A \equiv \oh(\d_{\m\a} \d_{\n\b} - \d_{\m\b} \d_{\n\a})
\eeq
and $P^{(2)}$, $P^{(1)}$, $P^{(0s)}$, $P^{(0w)}$ are spin projection
operators introduced by van Nieuwenhuizen \cite{vanN}. Using the
heat-kernel prescription

\bea
    lndet[M] &=& - \int^{\infty}_{\k^2/\L^4} {ds \over s}
\int d^5x \sqrt{{\overline g}} Tr[K(x,x,s)]
\non \\
     {d \over ds} K &=& - M K
\label{heatkern}
\eea
a short computation finds that

\beq
      c_0 = {21 \over 12 \sqrt{\pi}}
\left( {\Gamma(\oq) \over 4\pi} \right)^4 \L^6
+ {5 \over 2 \sqrt{\pi}} \d^4(0) \L^2 = c \L^6
\eeq
where the delta-function is also understood to be regulated by the
short-distance cutoff, i.e. $\d^4(0) \approx \L^4$.  The precise
value of $c$ is not really important; all we need is the fact that
it is positive.  Note that gauge-fixing is unnecessary in this
calculation, since $S_5$ is only invariant under $t_5$-independent
transformations.

   For finite $\l$ we have

\beq
      \int d^5x \sqrt{{\overline g}} = {3 T \over 8 G^2 \l^2}
\eeq
where $T$ is the extension of the $t_5$ axis (note that $t_5$ has
units of length squared), and therefore, to
leading order in $\L$ and $1/\l$,

\beq
       P(\l) \approx exp\left[ {3 c \L^6 T \over 8 G^2 \l^2} \right]
\label{result}
\eeq
This expression has been derived for $\l>0$.  For $\l<0$ there
are other solutions of the Einstein equations, representing
compact manifolds of non-spherical topology, again with
4-volume $\propto 1/\l^2$ \cite{Foam}.\footnote{There
are also noncompact solutions for $\l<0$; however (by assumption)
the functional integral is restricted to metrics representing compact
4-manifolds.}  A one-loop calculation
of $P(\l<0)$ around such backgrounds will produce an
expression similar to \rf{result}, differing only by replacement
of the factor $3/8$ in the exponent by some other positive constant.

   Equation \rf{result} is the promised result.
As in the Baum-Hawking expression \rf{BH}, there is clearly an
infinite peaking in the probability distribution of the regulated theory
at $\l \ra 0$, which is offered here as an explanation for
the vanishing of the cosmological constant.  Unlike the Baum-Hawking
expression \rf{BH}, eq. \rf{result} is a one-loop result, obtained from a
well-defined functional integral with a gravitational action which is
bounded from below.

\end{document}